\documentclass[twocolumn,preprintnumbers,floats,amsmath,amssymb,aps,prb,showpacs]{revtex4}

\usepackage{graphics}
\usepackage{epsfig}
\usepackage{subfigure}
\usepackage{amsmath}
\usepackage{dcolumn}
\usepackage{citesort}
\begin{document}

\preprint{(submitted to Phys. Rev. B)}

\title{Density-functional theory investigation of oxygen adsorption\\
at Pd(11$N$)($N$=3,5,7) vicinal surfaces}

\author{Yongsheng Zhang}
\author{Jutta Rogal}
\author{Karsten Reuter}%
\email{reuter@fhi-berlin.mpg.de}
\affiliation{
Fritz-Haber-Institut der Max-Planck-Gesellschaft,
Faradayweg 4-6, D-14195, Berlin, Germany 
}

\received{30th May 2006}

\begin{abstract}
We present a density-functional theory study addressing the on-surface adsorption of oxygen at the Pd(11$N$) ($N =3,5,7$) vicinal surfaces, which exhibit (111) steps and (100) terraces of increasing width. We find the binding to be predominantly governed by the local coordination at the adsorption site. This leads to very similar bonding properties at the threefold step sites of all three vicinal surfaces, while the binding at the central fourfold hollow site in the four atomic row terrace of Pd(117) is already very little disturbed by the presence of the neighboring steps. 
\end{abstract}



\maketitle

\section{Introduction}

Important materials processes and functions like oxide formation or oxidation catalysis have motivated a wealth of Surface Science studies on low Miller index surfaces of late transition metals (TMs) and their interaction with oxygen.\cite{reuter06} In recent years, experimental work has been increasingly complemented by independent information from first-principles theoretical studies, which appropriately account for the band structure of the TMs through supercell setups with periodic boundary conditions.\cite{payne92} En route towards an atomic-scale understanding of real TM surfaces, investigating the role of defects like steps, kinks or vacancies with similar rigor and detail as established for the low-index surfaces becomes a key endeavor. With respect to catalysis and oxide formation, such defects are frequently discussed as playing a prominent role, e.g. in form of active sites facilitating molecular dissociation or as nucleation centers.\cite{reuter06,gross03,handbook} 

Steps and kinks are suitably studied at high-index vicinal surfaces, which contain regular arrays of steps of specific orientation, separated by terraces of specific width. The advantage of these surfaces is that they (ideally) only contain one defined type of step in always the same local environment, and are amenable to periodic boundary supercell calculations. The disadvantage is that the extracted step properties can be masked by step-step interactions at decreasing terrace widths. This holds in particular, since one would precisely try to perform first-principles calculations addressing a certain step type by employing vicinal surfaces with a minimum terrace width, in order to reduce the computational burden. In this respect, we find it useful to conduct a systematic study of a family of vicinal surfaces that always exhibit the same step type, but terraces of increasing width. Examining the geometric and electronic structure of the clean and oxygen-covered surfaces, we can then extract by how much the local properties at a given step are affected by the presence of the neighboring steps.

\begin{figure}
\scalebox{0.43}{\includegraphics{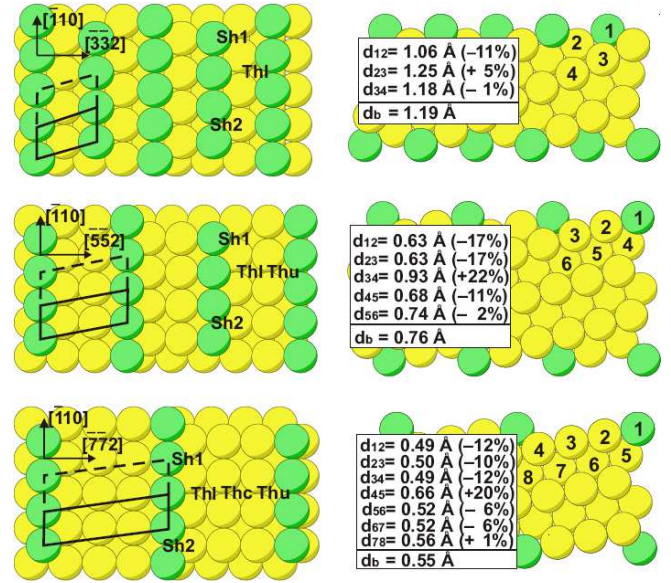}}
\caption{
(Color online) Top and side views of the atomic geometry of the family of Pd(11$N$) surfaces ($N$=3 [top panel], 5 [middle panel], 7 [bottom panel]) surfaces. Light grey (yellow) spheres represent Pd atoms, while the Pd step atoms are shown in dark grey (green), to better illustrate the step structure in the top view. Additionally shown in the top view are the $(1 \times 1)$ and $(1 \times 2)$ surface unit-cells (solid and dashed lines, respectively), as well as the high-symmetry sites for oxygen adsorption considered in this study (see text). The side view includes also the layer numbering and the relaxed spacings $d_{ij}$ between layers $i$ and $j$ (and in parenthesis the percent change $\Delta d_{ij}$ with respect to the bulk interlayer spacing $d_b$).}
\label{fig1} 
\end{figure}

Specifically, we use density-functional theory (DFT) to study the series of vicinal Pd(11$N$) ($N$=3,5,7) surfaces, additionally comparing them to the low-index Pd(100) and Pd(111) surfaces. Figure \ref{fig1} explains the geometric structure of the vicinals, which exhibit (111) steps and (100) terraces with 2, 3 and 4 atomic rows for Pd(113), Pd(115) and Pd(117), respectively. Besides the surface energetics and geometric structure with and without oxygen adsorbates, we also study the local density of states (LDOS) and initial-state surface core-level shifts (SCLS) as measures of the surface electronic structure. Apart from insight into details of the oxygen-metal chemical bond, our major conclusion is that the oxygen adsorption properties at the (111) step of Pd(117) (with 4 atomic row terraces) are already very little disturbed by the presence of the neighboring steps of the vicinal.

\section{Theory}

The DFT calculations were performed within the highly accurate full-potential augmented plane wave + local orbitals (LAPW/APW+lo) scheme \cite{wien2k}, using the generalized gradient approximation (GGA-PBE)\cite{perdew96} for the exchange-correlation (xc) functional. All surfaces were modeled by supercells, employing symmetric slabs (with O adsorption on both sides of the slab) and a minimum vacuum thickness of 12\,{\AA} to ensure the decoupling of consecutive slabs. The slab thickness was 7 layers for Pd(111) and Pd(100), 9 layers for Pd(113), 17 layers for Pd(115) and 19 layers for Pd(119), cf. Fig. \ref{fig1}. For the oxygen adsorption studies the thickness of the Pd(113), Pd(115) and Pd(117) slabs was increased to 13, 17, and 23, respectively. All structures were fully relaxed, keeping only the atomic positions in the central two slab layers at their fixed bulk positions.

The LAPW/APW+lo basis set parameters are as follows: $R_{\rm MT}^{\rm Pd} = 2.1$\,bohr, $R_{\rm MT}^{\rm O}= 1.1$\,bohr, wave function expansion inside the muffin tin spheres up to $l_{\rm max}^{\rm wf}= 12$, and potential expansion up to $l_{\rm max}^{\rm pot}= 6$. The energy cutoff for the plane wave representation in the interstitial region between the muffin tin spheres was $E_{\rm wf}^{\rm max} = 20$\,Ry for the wave functions and $E_{\rm pot}^{\rm max}=196$\,Ry for the potential. The core states are treated fully relativistic, while the semi-core and valence states are treated within a scalar relativistic approximation, i.e. spin-orbit coupling is included (neglected) for the core (semi-core and valence) states. Monkhorst-Pack (MP) grids were used for the Brillouin zone integrations. Specifically, we used $(6 \times 10 \times 1)$, $(4 \times 10 \times 1)$ and $(3 \times 10 \times 1)$ grids for the calculations of $(1 \times 1)$ surface unit-cells of Pd(113), Pd(115), and Pd(117), respectively. The selected calculations on Pd(100) and Pd(111) used for the comparison with low-index surfaces were done with a $(12 \times 12 \times 1)$ grid. For the larger surface cells, the grids were reduced accordingly to keep the same sampling of the reciprocal space. 

The central energetic quantities evaluated in our study are the surface energy $\gamma$ for the clean surfaces, and the binding energy $E_b$ for the oxygen adsorption. The prior is defined as
\begin{equation}
\gamma \;=\; \frac{1}{2A} \; \left( E_{\rm slab}- N_s E_{\rm bulk} \right)
\quad ,
 \label{surfeng}
\end{equation}
where $E_{\rm slab}$ and $E_{\rm bulk}$ are the total energy of the slab and of a bulk atom, respectively. $N_s$ is the number of atoms in the slab in the employed supercell, $A$ the corresponding surface area, and the factor of $\frac{1}{2}$ is used to account for the two equivalent surfaces of the slabs. In order to minimize the error due to the finite ${\bf k}$-point sampling, $E_{\rm bulk}$ is obtained for each surface from a calculation in exactly the same supercell as used for $E_{\rm slab}$, and filling the vacuum region with additional Pd layers to generate a bulk system. The oxygen binding energy is defined as
\begin{equation}
E_b \;=\; - \frac{1}{2 N_{\rm O}} \; \left( E_{\rm O/Pd} - E_{\rm Pd} - \frac{N_{\rm O}}{2} E_{\rm O_2} \right) \quad ,
\label{bindeng}
\end{equation}
where $N_{\rm O}$ is the number of O atoms present in the employed supercell, and $E_{\rm O/Pd}$, $E_{\rm Pd}$ and $E_{\rm O_2}$ are the total energies of the slab containing the oxygen adsorbate, the corresponding clean Pd slab, and the total energy of an isolated molecule, respectively. Again, the factor $\frac{1}{2}$ accounts for the fact that we adsorb oxygen at both sides of the symmetric slab.

With this definition, a positive $E_b$ value indicates that dissociative adsorption of O$_2$ is exothermic at $T = 0$\,K. To obtain the total energy of the isolated O$_2$ molecule, we exploit the relation $E_{\rm O_2} = 2 E_{\rm O}^{\rm atom} - D$, where $E_{\rm O}^{\rm atom}$ is the total energy of an isolated oxygen atom, and $D$ the theoretical O$_2$ binding energy \cite{ganduglia99,reuter02}. The isolated O atom is then calculated spin-polarized, inside a rectangular cell of side lengths $12 \times 13 \times 14$\,bohr, $\Gamma$-point sampling of the Brillouin zone and without spherically averaging the electron density in the open valence shell. For $D$ we employ the previously published ultra-converged value of $6.21$\,eV \cite{kiejna06}. Compared to the experimental binding energy of $5.12$\,eV \cite{curtiss91}, this value shows the well-known, substantial GGA-induced error. We note, however, that this will neither affect our discussion, nor our conclusions, since we focus on {\em differences} between oxygen binding energies at different surfaces and different sites, where this error cancels out.

Limitations in the numerical accuracy of our results come predominantly from the finite LAPW/APW+lo basis set and the supercell geometry. We performed extensive test calculations to estimate the importance of the various numerical approximations. Our tests concentrated on the convergence of the crucial energetic quantities of our study, using the surface energies of all five surfaces, as well as the oxygen binding energy in the Thu and Sh2 site (see definition below) in $(1 \times 1)$ unit-cells on the vicinals as representatives. Naming here only the factors that are most critical for the convergence, we increased the employed cutoff for the plane-wave basis set systematically from 20 to 30\,Ry, doubled the k-mesh density, and tested slab thicknesses with up to twice as many layers. From all these tests we estimate that our reported surface energies are converged to within 2\,meV/{\AA}$^2$, and the O binding energies to within 150\,meV/O atom. However, much more important for our study comparing the properties at different surfaces and sites are relative binding energy and surface energy {\em differences}. For these differences, a large part of the uncertainty, e.g. due to the description of the isolated O atom, cancels out, and our tests show that the numerical uncertainty for $\Delta E_b$ of O atoms at different sites or at different surfaces is $\pm 50$\,meV/O atom, and the relative surface energy differences are converged to within 0.5\,meV/{\AA}$^2$. As will become apparent below, this does not affect our discussion and conclusions.

\section{Clean vicinal surfaces}

\subsection{Geometric structure}

Using the Murnaghan equation of state, we compute the DFT-GGA-PBE lattice constant for Pd as $a = 3.947$\,{\AA} (neglecting zero-point vibrations) and the bulk modulus as $B = 157$\,GPa. These results are in excellent agreement with our previous study using the earlier WIEN97 LAPW code, which gave $a = 3.944$\,{\AA} and $B=163$\,GPa, respectively \cite{todorova04}. The slight overestimation of the lattice constant (about +2\% compared to the experimental value of $a_{\rm exp} = 3.89$\,{\AA} \cite{kittel}) and the (corresponding) underestimation of the bulk modulus ($B_{\rm exp}=181$\,GPa \cite{kittel}) are in line with analogous studies for other late $4d$ TMs.\cite{ganduglia99,reuter02,li02,silva05} The obtained surface relaxation pattern for the low-index Pd(100) and Pd(111) surfaces is also as expected in view of the trend understanding over the late $4d$ TMs e.g. discussed by Methfessel, Hennig and Scheffler \cite{methfessel92}. For both surfaces we obtain only insignificant ($< 1$\,\%) variations in the interlayer spacings of the topmost layers with respect to the bulk value.

Turning to the surface relaxation for the three vicinal surfaces, our results are summarized aside with the geometry side views in Fig. \ref{fig1}. One can clearly discern that the largest deviations from the bulk spacing are concentrated to those layers, in which the atoms have a lower coordination than in the bulk. When performing our slab thickness tests, we also fully relaxed slabs with up to twice as many layers as those shown in Fig. \ref{fig1}. However, no changes in the interlayer spacings were observed compared to the values listed in Fig. \ref{fig1}, and the deeper layers showed insignificant relaxations. Focusing therefore on the larger relaxations in the topmost layers, we notice that the signs of the percent changes $\Delta d_{ij}$ of the interlayer spacings compared to the bulk spacing exhibit a certain pattern. Starting with the topmost layer distance, we namely have $- + -$ for Pd(113), $- - + -$ for Pd(115) and $- - - + -$ for Pd(117), where ``$-$'' corresponds to a contraction and ``$+$'' to an expansion of the interlayer spacing.

\begin{table}
\caption{\label{tab1}
Comparison of the multilayer relaxation sequence of the Pd(11$N$), ($N=3,5,7$) vicinals to the key quantities of the atom-rows model \cite{tian00} and NN coordination model \cite{sun04}. A ``$-$'' in the relaxation sequence stands for a contraction of the corresponding interlayer spacing, and a ``$+$'' for an expansion. $n_{\rm row}$ is the number of atomic rows on the terraces and $n_{\rm NN}$ the number of undercoordinated layers at the surface. For comparison, the coordination is 12 in the fcc bulk, and 8 for a (100) terrace atom.}
\begin{ruledtabular}
\begin{center}
\begin{tabular}{ccccc} 
Vicinal & ~~$n_{\rm row}$~~ & NN Coordination & $n_{\rm NN}$  & Relaxation  \\
Surface &   & For Topmost Layers &   & Sequence \\ \hline 
Pd(113) & 2 & 7, 10, 12, 12, ... & 2 & $-+-$ \\
Pd(115) & 3 & 7, 8, 10, 12, 12, ...& 3 & $--+-$ \\
Pd(117) & 4 & 7, 8, 8, 10, 12, 12, ... & 4 & $---+-$ \\ 
\end{tabular}
\end{center}
\end{ruledtabular}
\end{table}

There are two theoretical models that explain these sequences as a consequence of Smoluchowski's ideas of charge smoothing (or equivalently, in a chemical language, through the bond-order bond-length correlation concept) \cite{smoluchowski41,pauling40,silva04}: The {\em atom rows} model \cite{tian00} predicts that for a surface with $n_{\rm row}$ atomic rows on the terraces, the first $(n_{\rm row} -1)$ topmost interlayer spacings exhibit a contraction, while the $n_{\rm row}$th one expands. The nearest-neighbor (NN) coordination model \cite{sun04} focuses instead on those surface layers which have a NN coordination smaller than the bulk. For a surface where the $n_{\rm NN}$ topmost layers are undercoordinated, the topmost $(n_{\rm NN} - 1)$ interlayer spacings are then expected to contract, while the $n_{\rm NN}$th one expands. As apparent from Table \ref{tab1} the multilayer relaxation sequences obtained for the present family of Pd(11$N$) vicinals conforms with both models, which is no surprise, since for the case of (111) steps the atom-rows model is actually a special case of the more general NN coordination concept.

\subsection{Energetics and electronic structure}

\begin{table}
\caption{\label{tab2} 
Computed surface energies of the low-index Pd(100) and Pd(111) surfaces, as well as of the three vicinal (Pd(11$N$), ($N=3,5,7$)) surfaces. Bulk-truncated refers to a bulk-truncated surface geometry, while relaxed corresponds to a fully relaxed surface geometry. We present all values in meV/{\AA}$^2$ and in eV/surface atom.}
\begin{ruledtabular}
\begin{tabular}{cc|ccc|c}
meV/{\AA}$^2$  &Pd(111)&Pd(113)&Pd(115)&Pd(117)&Pd(100) \\ \hline
Bulk-truncated &  87.9 & 100.3 &  99.4 &  99.1 &  96.4  \\
Relaxed        &  87.9 &  98.3 &  97.3 &  97.4 &  96.3  \\ \hline \hline
eV/atom        &Pd(111)&Pd(113)&Pd(115)&Pd(117)&Pd(100) \\ \hline
Bulk-truncated &  0.59 &  1.30 &  2.01 &  2.76 &  0.75  \\
Relaxed        &  0.59 &  1.27 &  1.97 &  2.71 &  0.75  \\
\end{tabular}
\end{ruledtabular}
\end{table}

Table \ref{tab2} summarizes the computed surface energies of the three vicinal surfaces and compares them to the surface energies of the two low-index surfaces Pd(100) and Pd(111). Due to the negligible layer relaxation of the latter two surfaces, a full relaxation of the surface geometry has only a negligible effect on their surface energy. As expected, this is different for the more open vicinal surfaces, where the atomic relaxation leads to some lowering. The values obtained for Pd(111) and Pd(100), 0.59 eV/atom and 0.75 eV/atom respectively, are in good agreement ($\pm 5$\,\%) with recent other DFT-GGA studies \cite{silva05,makkonen03,kwon05}. However, they are significantly different to two earlier trend studies, namely the linear-muffin-tin-orbital (LMTO) GGA work by Vitos {\em et al.} (Pd(111) 0.82 eV/atom; Pd(100) 1.15 eV/atom)\cite{vitos98}, and the LMTO work by Methfessel, Hennig and Scheffler employing the local-density approximation (LDA)\cite{perdew92} to the xc functional (Pd(111) 0.68 eV/atom; Pd(100) 0.89 eV/atom)\cite{methfessel92}. To assess the effect of the different approximate xc functional in the latter study, we recomputed the surface energies with the LDA and using the optimized LDA lattice constant (3.84\,{\AA}). The obtained values of 0.77\,eV/atom for Pd(111) and 1.02\,eV/atom for Pd(100) are again in good agreement with recent other LDA studies \cite{silva05,zabloudil06}. However, although the values are significantly different to the GGA numbers, they cannot resolve the discrepancy with respect to the early LDA work by Methfessel and coworkers. Instead, it seems to be the non-relativistic treatment of the valence and semi-core electrons in the latter work that is behind the difference. When deliberately switching off the scalar relativistic treatment employed in our work, we obtain values of 0.69\,eV/atom (Pd(111)) and 0.88\,eV/atom (Pd(100)) that are now very close to those obtained by Methfessel and coworkers. While we are therefore able to resolve the difference with respect to this work, we are unable to account for the discrepancy with respect to the TB-LMTO work by Vitos {\em et al.}. As already discussed by Da Silva, Stampfl and Scheffler \cite{silva05}, the reason might be that the atomic sphere approximation (ASA) employed by Vitos {\em et al.} does not allow for sufficient flexibility of electronic relaxation at the surface.

\begin{figure}
    \epsfig{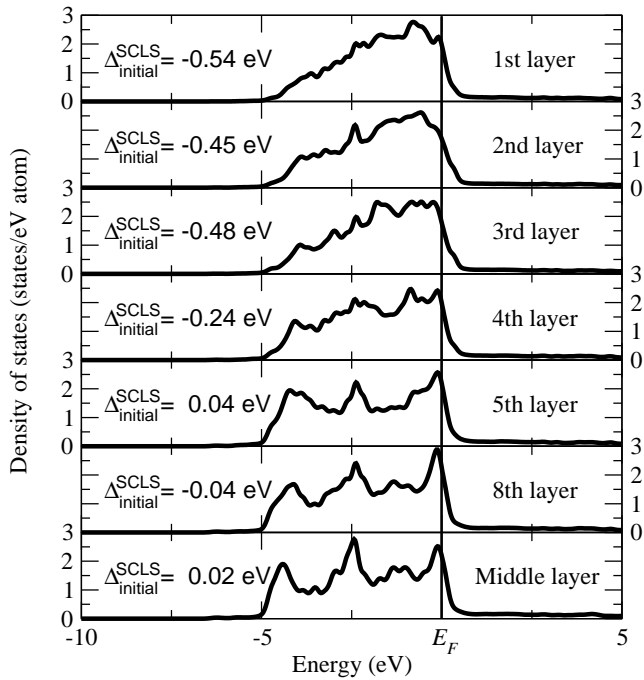}
\caption{\label{fig2} 
Local density of states (LDOS) in the different layers in Pd(117). Compared to the bulk-like LDOS in the central layer of the slab, the topmost four surface layers (corresponding to undercoordinated step and terrace atoms) exhibit a significant narrowing of the valence $4d$ band. The ensuing shift of the center of gravity of the valence $4d$ band induces a non-negligible shift $\Delta_{\rm initial}^{\rm SCLS}$ in the 3$d$ core-level positions with respect to the value in a bulk Pd atom. Note the similarity of the $\Delta_{\rm initial}^{\rm SCLS}$ exhibited by the two terrace atoms in Pd(117), 2nd and 3rd row, to the one of the topmost layer atoms in an infinite Pd(100) terrace, $\Delta_{\rm initial}^{\rm SCLS} = -0.50$\,eV.}
\end{figure}

As for the vicinal surfaces, the sequence of Pd(113), Pd(115) and Pd(117) surface energies gives already a first hint as to the influence of neighboring steps. In this respect, the obtained rather similar values in meV/{\AA}$^2$, cf. Table \ref{tab2}, suggest that step-step interactions are rather small despite the small terrace widths of these surfaces. This finding is in full agreement with a preceding tight-binding study by Raouafi {\em et al.}, which also determined negligible step-step interactions at $p(100)\times(111)$ vicinals of Rh, Pd and Cu \cite{raouafi02}. The fast decay of the local perturbation introduced by the (111) step is also nicely discerned in the surface electronic structure. Figure \ref{fig2} shows the local density of states (LDOS) calculated inside the muffin tin spheres of the atoms in the various surface layers at Pd(117). Compared to the bulk-like LDOS of the central layer in the slab, only the topmost four layers corresponding to step or terrace atoms with reduced NN coordination show a significant variation, predominantly in form of a narrowing of the valence $4d$ band. In order to maintain local charge neutrality, the center of gravity of the more than half full $d$ band has correspondingly moved up compared to the situation in a bulk Pd atom, which went hand in hand with a repulsive contribution to the Kohn-Sham potential \cite{lizzit01}. This potential change is felt by the core electrons, as well, and induces a shift in the position of the core-levels. We define this initial-state surface core-level shift (SCLS) with respect to a bulk atom as
\begin{equation}
\Delta_{\rm initial}^{\rm SCLS} \;=\; - \left[ \epsilon_{c}^{\rm surface} - \epsilon_{c}^{\rm bulk} \right] \quad ,
\end{equation}
where $\epsilon_{c}^{\rm surface}$ and $\epsilon_{c}^{\rm bulk}$ are the Kohn-Sham eigenvalues of the core state $c$ in a surface and a bulk atom. With this definition, the valence $d$ band narrowing at the surface leads to a negative shift, and we focus in this work in particular on the SCLS exhibited by the $3d$ Pd levels. 

\begin{table}
\begin{ruledtabular}
\begin{center}
\caption{\label{tab3}
Initial-state surface core-level shifts $\Delta^{\rm SCLS}_{\rm initial}$ for the topmost layers with reduced NN coordination, cf. Table \ref{tab1}, at the low-index and vicinal surfaces. All values are given in eV.}
\begin{tabular}{cccccc} 
$\Delta^{\rm SCLS}_{\rm initial}$& Pd(111)  & Pd(113) & Pd(115) & Pd(117) & Pd(100) \\ \hline
1st & -0.39    & -0.56   & -0.50   & -0.54   & -0.50   \\ 
2nd &          & -0.19   & -0.43   & -0.45   &         \\ 
3rd &          &         & -0.23   & -0.48   &         \\ 
4th &          &         &         & -0.24   &         \\ 
\end{tabular}
\end{center}
\end{ruledtabular}
\end{table}

$\Delta_{\rm initial}^{\rm SCLS}$ is one contribution to the measurable SCLS in x-ray photoemission spectroscopy (XPS), which in addition comprises also the screening contribution of the valence electrons in response to the created core hole \cite{lizzit01}. However, it is not so much this connection to an experimentally accessible quantity that makes us interested in the initial-state SCLS, but rather that the $\Delta_{\rm initial}^{\rm SCLS}$ are a very sensitive probe of changes in the local electronic structure of an atom in different environments. In this respect we note that the $\Delta_{\rm initial}^{\rm SCLS}$ exhibited by the eightfold coordinated 2nd and 3rd layer terrace atoms at Pd(117), cf. Fig. \ref{fig2}, are already quite similar to the one we compute for a topmost layer atom in an infinite Pd(100) terrace, $-0.50$\,eV/atom. In contrast, all deeper layers with 12fold bulk-like coordination exhibit only insignificant shifts. Table \ref{tab3} shows that the nearest-neighbor coordination is in fact the overriding parameter determining the initial-state SCLSs at all surfaces studied here. Comparing to the NN coordination listed in Table \ref{tab1}, one even finds a roughly linear relation of the $\Delta_{\rm initial}^{\rm SCLS}$ with the number of NN Pd atoms, which indicates that perturbations further away than the immediate NN shell are rapidly screened away.

\section{Oxygen adsorption at Pd(11$N$) vicinal surfaces}

\subsection{Binding energy}

\begin{table}
\begin{ruledtabular}
\begin{center}
\caption{\label{tab4} 
Binding energies of oxygen atoms at the different highly-coordinated terrace and step sites of the three Pd(11$N$) ($N$=3,5,7) surfaces, see Fig. \ref{fig1} for an illustration of all sites and the employed nomenclature. Listed are results for the adsorption of one oxygen atom per $(1 \times 1)$ and $(1 \times 2)$ surface unit-cell. At the Pd(113) surface, the Thu and Thl sites coincide, which is why only one value is given. All values are in eV.}
\vspace{0.5 cm}
\begin{tabular}{cccccc} 
\hline
$(1 \times 1)$ & Sh1 & Sh2 & Thu & Thl & Thc  \\ \hline
Pd(113)        & 0.82& 0.87&\multicolumn{2}{c}{ 0.60}&      \\ 
Pd(115)        & 0.80& 0.89& 0.84& 0.52&      \\ 
Pd(117)        & 0.83& 0.94& 0.86& 0.57& 0.80 \\ 
\hline
$(1 \times 2)$ & Sh1 & Sh2 & Thu & Thl & Thc  \\ \hline
Pd(113)        & 1.08& 1.17&\multicolumn{2}{c}{ 0.96}&      \\ 
Pd(115)        & 1.04& 1.17& 1.27& 0.92&      \\ 
Pd(117)        & 1.05& 1.22& 1.27& 0.98& 1.25 \\ 
\end{tabular}
\end{center}
\end{ruledtabular}
\end{table}

Interested in evaluating the effect of the nearby steps in the series of Pd(11$N$) surfaces, we consider the adsorption of oxygen as a probe of the local chemical properties. For this we focus on the binding of atomic O to the highly-coordinated high-symmetry sites offered by the vicinal surfaces, namely the fourfold coordinated hollow sites on the terraces and the threefold coordinated hollow sites at the steps, cf. Fig. \ref{fig1}. This yields two step sites at each vicinal surface, the Sh1 site where the oxygen atom is coordinated to one step atom and the Sh2 site where the oxygen atom is coordinated to two step atoms. At the Pd(117) surface, there are three terrace hollow sites, one at the lower edge of the step (Thl), one at the upper edge of the step (Thu) and one in the center of the terrace (Thc). Due to the reduced terrace width, there are only Thl and Thu sites at Pd(115), and at Pd(113) the lower and upper edge terrace hollow sites coincide. To analyze the effect of coverage we calculated the binding energy of one oxygen atom at any of these sites both in $(1 \times 1)$ and $(1 \times 2)$ surface unit cells. In the prior, all sites of one type are then covered, e.g. all Sh1 sites along a step, while in the latter occupied sites alternate with empty ones, e.g. one Sh1 site covered, the neighboring one empty etc. 

The obtained binding energies are compiled in Table \ref{tab4} and can be understood in a rather simple and local picture. Central for the stability at a given site are the number of directly coordinated surface Pd atoms and their respective Pd coordination ($n$fold coordinated terrace or step atoms). Sites which are identical with respect to these criteria exhibit binding energies that are almost degenerate within our numerical accuracy ($\pm 50$\,meV/O atom). Prominent examples are the similar binding energies at the Sh1 site at all three surfaces (binding to one 7fold coordinated step and two 10fold coordinated terrace atoms) and at the Sh2 site at all three surfaces (binding to two 7fold coordinated step and one 10fold coordinated terrace atom). Due to the narrow terrace widths this equivalence is only established for the Thu (and separately for the Thl) sites at the Pd(115) and Pd(117) surfaces. There, Thl corresponds to binding to two 8fold coordinated and two 10fold coordinated terrace atoms, while Thu corresponds to binding to two 7fold coordinated step atoms and two 8fold coordinated terrace atoms. The binding site Thc in the center of a terrace (coordination to four 8fold terrace atoms) develops only at the Pd(117) surface, but is then already comparable to the fourfold hollow site at Pd(100), corresponding to an infinite terrace width. We arrive at this assessment from calculating $(3 \times 1)$ and $(4 \times 1)$ overlayers at Pd(100) with one oxygen atom per surface unit-cell. The obtained binding energies are 0.84\,eV/O atom and 0.85\,eV/O atom, respectively, which shows that the lateral interactions between the neighboring O atoms along the long axis of the unit-cell are already negligible. On the other hand, these binding energies are very close to the one computed for the Thc site at Pd(117), 0.80\,eV/O atom, revealing an only marginal influence of the neighboring steps in the latter case. Furthermore, the predominant role of the local coordination seems to hold independently of the coverage, since our calculations in $(1 \times 2)$ surface unit-cells, cf. Table \ref{tab4}, show exactly the same pattern. In this case, we compute an O binding energy of 1.29\,eV/O atom in a $(3 \times 2)$ overlayer on Pd(100), which is again very similar to the value of 1.25\,eV/O atom at the Thc site.

Also the stability ordering among the different site types largely follows a local coordination picture, when recognizing that the bonding is stronger to lower-coordinated Pd atoms than to higher-coordinated ones. Correspondingly, the Sh2 site is more stable than the Sh1 (bonding to two step atoms vs. bonding to one step atom), while the ordering among the terrace sites goes with decreasing stability as Thu, Thc and Thl. Interestingly, the most stable terrace site (Thu) is slightly more stable than the most stable step site (Sh2) at the lower $(1 \times 2)$ coverage studied, while this order is reversed at the higher $(1 \times 1)$ coverage. This might again be explained in a local coordination picture, when realizing that in the $(1 \times 1)$ arrangement, neighboring O atoms in rows of any of the terrace sites share two Pd atoms, while in rows of either of the two step sites they share only one Pd atom. In terms of a bond-order conservation picture, it is then intelligible that equivalent terrace sites are $\sim 0.4$\,eV/O atom less stable in the denser $(1 \times 1)$ overlayer compared to the $(1 \times 2)$ one, while for the step sites this reduction is only $\sim 0.25$\,eV/O atom. With the Thu site only slightly more favorable than the Sh2 site in the $(1 \times 2)$ arrangement, the smaller reduction for the Sh2 site when going to the dense $(1 \times 1)$ arrangement leads then to slight preference for the Sh2 site. We speculate that this change in the stability ordering of terrace and step sites with coverage may lead to interesting ordering phenomena. Addressing this topic requires a proper evaluation of the partition function, on which we focus in ongoing work.

\subsection{Geometric and electronic structure}

\begin{table*}
\begin{ruledtabular}
\caption{\label{tab5}
O-Pd bond lengths (in {\AA}) at the different sites in $(1 \times 1)$ overlayers. The shorter bond lengths are always to the lower coordinated Pd atoms, cf. Fig. \ref{fig1} and Table \ref{tab2}. At the Pd(113) surface, the Thu and Thl sites coincide, which is why only one set of values is given.}
\begin{tabular}{c|ccccc}
       & Sh1            & Sh2            & Thu                 & Thl & Thc \\ \hline
Pd(113)& 1.97/2.02/2.02 & 1.98/1.98/2.02 & \multicolumn{2}{c}{2.00/2.00/2.47/2.47} & \\ 
Pd(115)& 1.99/2.01/2.01 & 1.99/1.99/2.03 & 2.15/2.15/2.14/2.14 & 2.05/2.05/2.30/2.30 &\\ 
Pd(117)& 1.98/2.01/2.01 & 1.98/1.98/2.03 & 2.15/2.15/2.14/2.14 & 2.05/2.05/2.30/2.30 & 2.13/2.13/2.16/2.16 \\ 
\end{tabular}
\end{ruledtabular}
\end{table*}

The calculated relaxed surface geometries reflect the local coordination picture developed in the previous section. Focusing first on the O-Pd bond lengths, the data for the $(1 \times 1)$ O overlayers compiled in Table \ref{tab5} clearly reveals two trends (equivalent results are obtained for the $(1 \times 2$) O overlayers): Longer O-Pd bond lengths are found at higher coordinated adsorption sites (fourfold terrace sites vs. threefold step sites), and shorter bond lengths result in bonds to lower coordinated Pd atoms (7fold coordinated step atoms vs. 8 or 10fold coordinated terrace atoms). Equivalent sites at the different surfaces that possess the same coordination and hence similar binding energies as discussed above, also exhibit similar bond lengths. The local geometry at the Sh1 and Sh2 sites is thus the same at all three vicinials, and so is the one at the Thu and Thl sites at Pd(115) and Pd(117). Since the bonds to lower coordinated Pd atoms are shorter, the threefold step sites Sh1 and Sh2 are not exactly threefold, but a slightly shorter bond length results to the step atoms. The same occurs for the terrace sites, where the inequivalence of the Thl site at the bottom of the step is most pronounced: Here, the O atom is significantly shifted away from ideal fourfold site, away from the step and towards the terrace center. Just as with respect to the binding energies, the Thc site is already very much comparable in its local geometry to the fourfold hollow site at a similarly covered Pd(100) surface: For $(3 \times 1)$ and $(4 \times 1)$ overlayers with one O atom per surface unit-cell we compute bondlengths of 2.14\,{\AA}, which are almost identical to the 2.13\,{\AA}/2.16\,{\AA} found at Pd(117). Interestingly, this bond length similarity to the low-index surfaces holds even for the threefold step sites Sh1 and Sh2, which exhibit bond lengths that are very close to the $\approx 2$\,{\AA} computed for the threefold hollow sites in various overlayers on Pd(111) \cite{todorova04}. 

\begin{figure}
\scalebox{0.42}{\includegraphics{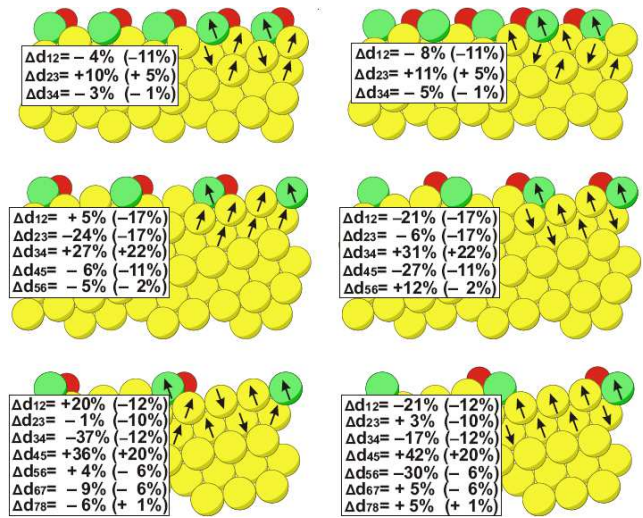}}
\caption{\label{fig3}
(Color online) Geometry side views for adsorption in the Sh2 (left panels) and Thu (right panels) sites in $(1 \times 1)$ overlayers at the Pd(113) [top panels], Pd(115) [middle panels], and Pd(117) [bottom panels]. The insets show the percent change of the layer spacings with respect to the bulk spacing $\Delta d_{ij}$. For comparison, the value in parenthesis gives the corresponding $\Delta d_{ij}$ of the clean surface. The arrows indicate the direction of the substrate atom displacement (including the lateral relaxation) with respect to the position in the clean surface. Light grey (yellow) spheres represent Pd atoms, dark grey (green) spheres Pd step atoms, and small dark (red) spheres O atoms.}
\end{figure}

\begin{table*}
\begin{ruledtabular}
\caption{\label{tab6}
Surface relaxation pattern of the Pd(11$N$) ($N=3,5,7$) vicinal surfaces, both for the clean surfaces and with a $(1 \times 1)$ O overlayer adsorbed. Shown are the interlayer spacings $d_{ij}$ (see Fig. \ref{fig1} for a definition of the layer numbers), as well as in parenthesis the percent change $\Delta d_{ij}$ with respect to the interlayer spacing $d_b$ in the bulk. See Fig. \ref{fig1} for an illustration of the various adsorption sites.}
\begin{tabular}{lcccc}
Pd(113)  & Sh1 & Sh2 & Thu/Thl & clean \\ \hline
$d_{12}$ \,($\Delta d_{12}$) & 1.09\,{\AA} \; ($-$9\%) & 1.15\,{\AA} \; ($-$4\%) & 1.10\,{\AA}  \; ($-$8\%) & 1.06\,{\AA}    ($-$11\%) \\
$d_{23}$ \,($\Delta d_{23}$) & 1.32\,{\AA}     (+11\%) & 1.31\,{\AA}     (+10\%) & 1.32\,{\AA} (+11\%)  & 1.25\,{\AA} \; (+5\%) \\
$d_{34}$ \,($\Delta d_{34}$) & 1.11\,{\AA} \; ($-$7\%) & 1.15\,{\AA} \; ($-$3\%) & 1.12\,{\AA}  \; ($-$5\%) & 1.18\,{\AA} \; ($-$1\%) \\
$d_{45}$ \,($\Delta d_{45}$) & 1.23\,{\AA} \; (+3\%)   & 1.21\,{\AA} \; (+2\%)   & 1.23\,{\AA}  \; (+3\%)   & 1.19\,{\AA} \;\,\;\,\; (0\%)\\
$d_{56}$ \,($\Delta d_{56}$) & 1.17\,{\AA} \; ($-$2\%) & 1.19\,{\AA} \;\,\;\,\; (0\%)  & 1.18\,{\AA} \; ($-$1\%) & 1.19\,{\AA} \;\,\;\,\; (0\%)
\end{tabular}
\vspace{0.1 cm}
\begin{tabular}{lccccc}
Pd(115)  & Sh1 & Sh2 & Thu & Thl & clean \\ \hline
$d_{12}$ \,($\Delta d_{12}$) & 0.74\,{\AA} \; ($-$2\%)& 0.79\,{\AA} \; (+5\%) & 0.60\,{\AA} ($-$21\%) & 0.64\,{\AA} ($-$15\%) & 0.63\,{\AA} ($-$17\%) \\
$d_{23}$ \,($\Delta d_{23}$) & 0.62\,{\AA} ($-$19\%)& 0.58\,{\AA} ($-$24\%) & 0.71\,{\AA} \;($-$6\%) & 0.73\,{\AA} \; ($-$4\%) & 0.63\,{\AA} ($-$17\%) \\
$d_{34}$ \,($\Delta d_{34}$) & 0.95\,{\AA} (+24\%)& 0.97\,{\AA} (+27\%) & 1.00\,{\AA} (+31\%) &  0.82\,{\AA} \; (+8\%) & 0.93\,{\AA} (+22\%) \\
$d_{45}$ \,($\Delta d_{45}$) & 0.70\,{\AA} \; ($-$7\%)& 0.71\,{\AA} \; ($-$6\%) & 0.55\,{\AA} ($-$27\%) & 0.85\,{\AA} (+11\%) & 0.68\,{\AA} ($-$11\%) \\
$d_{56}$ \,($\Delta d_{56}$) & 0.71\,{\AA} \; ($-$6\%)& 0.72\,{\AA} \; ($-$5\%) & 0.85\,{\AA} (+12\%) & 0.62\,{\AA} ($-$19\%) & 0.74\,{\AA} \; ($-$2\%) \\
$d_{67}$ \,($\Delta d_{67}$) & 0.83\,{\AA} \; (+9\%)& 0.81\,{\AA} \; (+7\%) & 0.77\,{\AA} \; (+1\%) & 0.81\,{\AA} \; (+6\%) & 0.79\,{\AA} \;  (+4\%) \\
$d_{78}$ \,($\Delta d_{78}$) & 0.72\,{\AA} \; ($-$5\%)& 0.74\,{\AA} \; ($-$2\%) & 0.74\,{\AA} \; ($-$2\%) & 0.80\,{\AA} \; (+5\%) & 0.75\,{\AA} \;  ($-$2\%)
\end{tabular}
\vspace{0.1 cm}
\begin{tabular}{lcccccc}
Pd(117)  & Sh1 & Sh2 & Thu & Thc & Thl & clean \\ \hline
$d_{12}$ \,($\Delta d_{12}$) & 0.66\,{\AA} (+20\%) & 0.66\,{\AA} (+20\%) & 0.44\,{\AA} ($-$21\%) & 0.48\,{\AA} ($-$13\%) & 0.48\,{\AA} ($-$12\%) & 0.49\,{\AA} ($-$12\%) \\
$d_{23}$ \,($\Delta d_{23}$) & 0.52\,{\AA} \; ($-$6\%) & 0.54\,{\AA} \; ($-$1\%) & 0.57\,{\AA} \; (+3\%) & 0.51\,{\AA} \; ($-$7\%) & 0.46\,{\AA} ($-$16\%) & 0.50\,{\AA} ($-$10\%) \\
$d_{34}$ \,($\Delta d_{34}$) & 0.29\,{\AA} ($-$47\%) & 0.35\,{\AA} ($-$37\%) & 0.46\,{\AA} ($-$17\%) & 0.51\,{\AA} \; ($-$7\%) & 0.51\,{\AA} \; ($-$8\%) & 0.49\,{\AA} ($-$12\%) \\
$d_{45}$ \,($\Delta d_{45}$) & 0.80\,{\AA} (+46\%) & 0.75\,{\AA} (+36\%) & 0.78\,{\AA} (+42\%) & 0.65\,{\AA} (+18\%) & 0.70\,{\AA} (+28\%) & 0.66\,{\AA} (+20\%) \\
$d_{56}$ \,($\Delta d_{56}$) & 0.59\,{\AA} \; (+8\%) & 0.57\,{\AA} \; (+4\%) & 0.39\,{\AA} ($-$30\%) & 0.64\,{\AA} (+17\%) & 0.49\,{\AA} ($-$11\%) & 0.52\,{\AA} \; ($-$6\%) \\
$d_{67}$ \,($\Delta d_{67}$) & 0.45\,{\AA} ($-$18\%) & 0.50\,{\AA} \; ($-$9\%) & 0.58\,{\AA} \; (+5\%) & 0.36\,{\AA} ($-$34\%) & 0.66\,{\AA} (+20\%) & 0.52\,{\AA} \; ($-$6\%) \\
$d_{78}$ \,($\Delta d_{78}$) & 0.49\,{\AA} ($-$11\%) & 0.52\,{\AA} \; ($-$6\%) & 0.58\,{\AA} \; (+5\%) & 0.63\,{\AA} (+15\%) & 0.40\,{\AA} ($-$28\%) & 0.56\,{\AA} \; (+1\%) \\
$d_{89}$ \,($\Delta d_{89}$) & 0.69\,{\AA} (+25\%) & 0.61\,{\AA} (+12\%) & 0.58\,{\AA} \; (+5\%) & 0.59\,{\AA} \; (+6\%) & 0.60\,{\AA} \; (+10\%) & 0.56\,{\AA} \; (+1\%) \\
\end{tabular}
\end{ruledtabular}
\end{table*}

The chemisorption of oxygen also induces substantial changes in the relaxation pattern of the underlying Pd substrate atoms. Similar to the situation at lower coverages at the low-index Pd(111) \cite{todorova04} and Pd(100) surfaces, the main effect of the adsorbed oxygen is to pull the directly coordinated Pd atoms out of the surface. However, while primarily at Pd(100) this leads simply to a significant expansion of the first layer spacing ($\Delta d_{12} = +12$\,\% at $(1 \times 1)$ coverage), the relaxation is more complicated at the more open vicinal surfaces, which also allow for quite some lateral displacements of the Pd atoms. The corresponding data is shown in Fig. \ref{fig3} for O adsorption in the most stable step and terrace site, Sh2 and Thu respectively, and compiled in Table \ref{tab6} for all adsorption sites. Due to the lateral displacements, the oxygen-induced changes in the relaxation pattern also propagate much further than just the nearest-neighbor Pd shell. Here, the locality is thus not that pronounced, but apparently this has no significant consequences on the afore discussed binding energies. 

\begin{figure*}
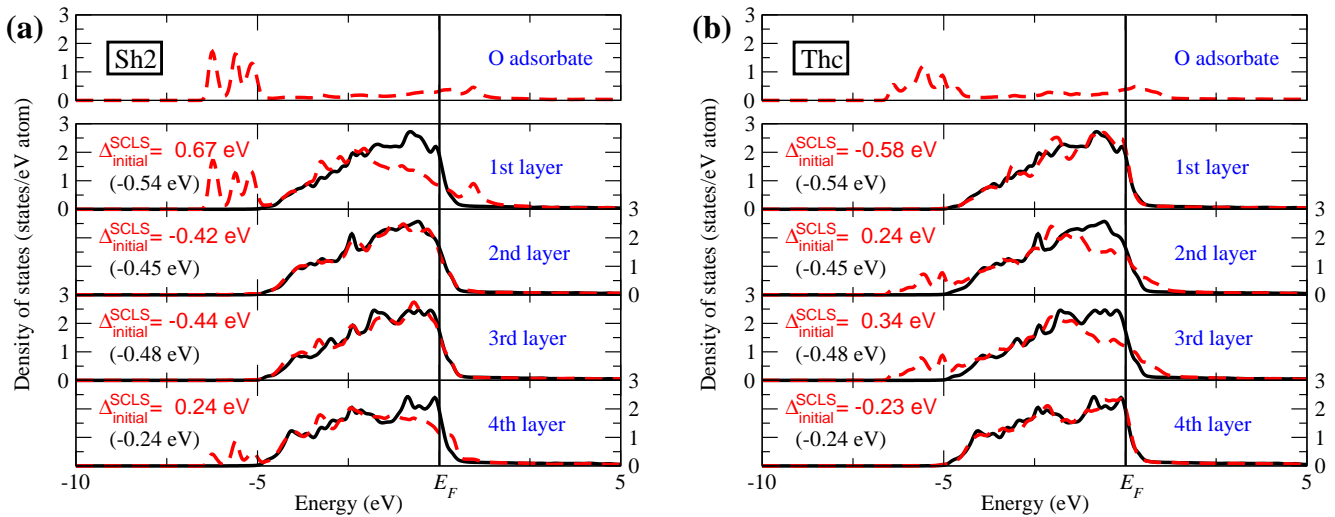

  \subfigure{ 
     \epsfig{bbllx=35,bblly=52,bburx=536,bbury=454,clip=,
             file=pics/fig4-1.eps,width=0.98\columnwidth}
	     } \quad
  \subfigure{ 
     \epsfig{bbllx=35,bblly=52,bburx=536,bbury=454,clip=,
             file=pics/fig4-2.eps,width=0.98\columnwidth}
	     } 
\caption{\label{fig4}
(Color online) Oxygen-induced changes in the local density of states (LDOS) and initial-state SCLSs. Shown are the data for oxygen adsorption into the Sh2 step (left panels) and Thc terrace site (right panels) at Pd(117). Dashed lines correspond to the LDOS with the O adsorbate (inside the O muffin tin in the top panel and inside the Pd muffin tin in all other panels), solid lines correspond to the clean surface. The $\Delta_{\rm initial}^{\rm SCLS}$ exhibited by the Pd terrace atoms directly coordinated to the O atom in the Thc site, layers 2 and 3, are very similar to the 0.32\,eV computed for the directly-coordinated Pd atoms in a $(3 \times 1)$ O-overlayer at Pd(100). The values in parenthesis give the initial-state SCLSs at the clean surface for comparison.}
\end{figure*}

The perturbation in the electronic system created by the chemisorbed oxygen atom is again very localized, as is nicely reflected in the layer-resolved LDOSs. Figure \ref{fig4} exemplifies this for oxygen at the Sh2 step and Thc terrace site at Pd(117). In both cases, the local density of states is primarily affected in the surface layers corresponding to the Pd atoms directly coordinated to the O adsorbate, namely layers 1 and 4 for the Sh2 site, and layers 2 and 3 for the Thc site, cf. Fig. \ref{fig1}. In these layers, we find O-Pd bonding states just below and antibonding states just above the nearly fully occupied Pd $4d$ band. In the shown LDOS for the $(1 \times 1)$ overlayer, the bonding states show some substructure due the presence of a noticeable oxygen-oxygen interaction along the rows in the short direction of the surface unit-cell. This leads to the formation of an adsorbate band structure as previously discussed for high-coverage O overlayers at the low-index Pd(111) surface \cite{todorova04}. The surface layers corresponding to atoms not belonging to the immediate NN shell of the oxygen adsorbate show only insignificant variations in the LDOS compared to the case of the clean surface, as does the LDOS in the deeper layers. Correspondingly, it is also only the directly coordinated surface layers that exhibit appreciable adsorbate-induced changes in the initial-state SCLSs, cf. Fig. \ref{fig4}. The widening of the $d$-band leads then to more positive shifts compared to the clean surface SCLSs \cite{lizzit01}. The stronger the formed O-Pd bond, the larger the disturbance of the local potential and the larger the ensuing shift. In this respect, we find O-induced shifts, i.e. a change of $\Delta^{\rm SCLS}_{\rm initial}$ compared to the value at the clean surface, of +1.21\,eV and +0.48\,eV at the 7fold and 10fold coordinated Pd atoms to which the O atom binds in the Sh2 site, and shifts of +0.69\,eV and +0.82\,eV at the two types of 8fold terrace Pd atoms to which the O atom binds in the Thc site. This indicates a significantly different bonding of the O atom to the different NN Pd atoms, which correlates with the varying O-Pd bond lengths shown in Table \ref{tab5} and suggests that the 7fold coordinated step atoms may offer a stronger bonding compared to the higher-coordinated terrace atoms. That the overall binding energy at the Sh2 site is nevertheless still comparable to the most stable terrace sites, cf. Table \ref{tab4}, results simply because the latter offer a higher (fourfold) coordination to Pd atoms compared to the threefold sites at the (111) step.

The perturbation due to oxygen adsorption at all other investigated sites is as localized as in the just discussed case of the Sh2 and Thc sites. In all cases, we only find significant O-induced shifts in the initial-state SCLS at the directly coordinated Pd atoms. Just as for the binding energies and O-Pd bond lengths, also these SCLSs are very similar for adsorption into equivalent sites at the three vicinal surfaces. In a $(3 \times 1)$ overlayer at Pd(100) we compute an O-induced SCLS shift of +0.82\,eV exhibited by the directly coordinated Pd atoms, which is again very similar to the above cited values for O adsorption into the Thc terrace site at a Pd(117) surface. This corroborates our understanding that the neighboring steps have already a negligible influence on the O bonding at this site in the center of an only four atomic row wide terrace.

\section{Conclusions}

We presented a DFT-GGA study addressing the on-surface adsorption of oxygen at Pd(11$N$) ($N$=3,5,7) vicinal surfaces. Apart from providing detailed insight into the oxygen-metal chemical bond, our particular interest was to investigate the influence on the adsorption properties by the neighboring steps that are separated by 2, 3 and 4 atomic row wide terraces at the three surfaces, respectively. The computed data on binding energies, local density of states and initial-state surface core-level shifts points at a very localized perturbation created by the chemisorbed oxygen atoms, primarily concentrating on the nearest neighbor Pd atoms. Central for the stability at a given site are then the number of directly coordinated surface Pd atoms (threefold hollow sites at the (111) steps vs. fourfold hollow sites at the (100) terraces) and their respective Pd coordination ($n$fold coordinated Pd terrace or step atoms). Sites which are equivalent with respect to these criteria exhibit binding energies that are degenerate to within 70\,meV/O atom. Examples are the step sites at all three vicinal surfaces, while the hollow site in the center of the four atomic row (100) terrace at Pd(117) is already very much comparable to the ones at the low-index Pd(100) surface. Interestingly, the stability at the threefold (111) step sites is very similar to the stability at the fourfold (100) terrace sites, and their energetic order depends on the local oxygen concentration. We speculate that this may lead to interesting ordering phenomena, on which we focus in ongoing work.

\begin{acknowledgments}
The EU is acknowledged for financial support under contract no. NMP3-CT-2003-505670 (NANO$_2$). 
\end{acknowledgments}

\end{document}